\documentclass[aps,prb,twocolumn,floatfix,floats,showpacs,superscriptaddress,raggedbottom,amsmath,amssymb]{revtex4}

\usepackage{graphicx}
\usepackage{dcolumn}
\usepackage{bm}
\usepackage{ulem}

\def\vec#1{{\bf#1}}
\def\eq#1{Eq.\ (\ref{#1})}

\def\fig#1{Fig.\ \ref{#1}}

\def\figs#1{Figs.\ \ref{#1}}

\begin{document}


\title{Engineered spin phase diagram of two interacting electrons\\
 in semiconductor nanowire quantum dots}

\author{Yan-Ting Chen}
\affiliation{Department of Electrophysics, National Chiao Tung
  University, Hsinchu 30010, Taiwan, Republic of China}
\author{Shun-Jen Cheng}
\email{sjcheng@mail.nctu.edu.tw}
\affiliation{Department of Electrophysics, National Chiao Tung
  University, Hsinchu 30010, Taiwan, Republic of China}
\author{Chi-Shung Tang}
\affiliation{Department of Mechanical Engineering, National United
University, Miaoli 36003, Taiwan, Republic of China}



\date{\today}

\begin{abstract}

Spin properties of two interacting electrons in a quantum dot (QD) embedded in a nanowire  with controlled aspect ratio and longitudinal magnetic fields
are investigated by using a configuration interaction (CI) method and exact diagonalization (ED) techniques. The developed CI theory
based on a three-dimensional (3D) parabolic model provides  explicit formulations of the Coulomb matrix elements and allows for straightforward
and efficient numerical implementation. Our studies reveal fruitful features of spin singlet-triplet transitions of two electrons confined in a nanowire quantum dot (NWQD),
as a consequence of the competing effects of geometry-controlled kinetic energy quantization, the various Coulomb interactions, and spin Zeeman energies.
The developed theory is further employed to study the spin phase diagram of two quantum-confined electrons in the regime of ``cross over'' dimensionality,
from quasi-two-dimensional (disk-like) QDs to finite one-dimensional (rod-like) QDs.

\end{abstract}

\maketitle

\section{Introduction}

Stimulated by recent success in coherent control of two-electron  spin
in laterally coupled quantum dots (QDs),~\cite{Petta05} the spin states
of two interacting electrons in semiconductor QDs have received
increasingly considerable attention. Accessible and engineerable spin
states of few electrons in QDs thus have become one of the basic
features required by the quantum information applications in which
electron spins are utilized as quantum bit.~\cite{Loss98,Loss09} For
two-dimensional (2D) epitaxial QDs, magnetic field induced spin
singlet-triplet (ST) transitions of two-electron ground states have
been studied extensively for
years.~\cite{Kouwenhoven97,Kouwenhoven01,Reimann02,Ellenberger06}  The
underlying physics of the ST transitions is usually associated with the energetic competition between quantized kinetic energies, the coulomb
interactions, and spin Zeeman energies.  Reversely switching the singlet and triplet spin states of a
lateral two-electron QD is feasible by utilizing electrical
control.~\cite{Kyriakidis02}  Moreover, it has been both theoretically
and experimentally shown that more complex oscillating spin phases can be
generated either by reducing the lateral confinement or by increasing
an applied magnetic field.~\cite{Wagner92,Hawrylak93-prl,Peeters99,Tarucha07}

Recently, the local-gate electrical
depletion~\cite{Fasth05,Ensslin06,Ensslin07} and the bottom-up grown
techniques~\cite{Bjork04,Bjork05}  have been developed for the fabrication
of few-electron QDs embedded in a nanowire.  These experimental
developments open up an opportunity of exploring the \textit{cross
over} mechanisms from the 2D (disk-like) to the finite 1D (rod-like) QD
regimes. Such nanowire quantum dots (NWQDs) are advantageous for
geometrical control over a wide rage of aspect ratio $a$ (typically
from $a\sim 10^{-1}$ to $a\gg 1$).~\cite{Bjork04,Bjork05}  The
excellent versatility of shape and dimensionality makes NWQDs
a suitable nanomaterial for scalable quantum
electronics.  Very recently, successful fabrication of single electron
transistors made of InAs based gate-defined NWQDs and observations of
the singlet-triplet transitions of two electrons in the QDs have been
demonstrated.~\cite{Fasth07} How the highly tunable longitudinal
confinement of NWQD affects and can be utilized to tailor the spin
properties of few electrons in NWQDs are interesting subjects worth
studying.

The above experimental efforts motivate us to perform a theoretical investigation of the spin states of two electrons in InAs-based NWQDs ~\cite{Fasth07}
by using a developed configuration interaction (CI) theory and exact diagonalization techniques.~\cite{Hawrylak03}
The developed CI theory is based on the 3D parabolic model with arbitrary transverse and longitudinal confinement strengths~\cite{Nazmitdinov97,Lin01}
and provides explicit generalized formulations of the Coulomb matrix, and thus allows for straightforward and efficient numerical or even semi-analytical implementation widely applicable
for various cylindrically symmetric QDs. Our exact diagonalization studies of two-electron charged NWQDs with controlled geometric aspect ratios
and longitudinal magnetic fields reveal fruitful features of spin singlet-triplet transitions, as a consequence of the competing effects of geometry-engineered
kinetic energy quantization, the various Coulomb interactions, and spin Zeeman energies. The developed theory is further employed to study the spin phase diagram of
two quantum-confined electrons in the regime of ``cross over'' dimensionality from quasi-2D (disk-like) QDs to finite 1D (rod-like) QDs.

This article is organized as follows: Section II describes the
theoretical model  and the developed configuration interaction theory
for few-electron problems of three-dimensionally confining quantum dots. In Sec. III, we present and discuss the calculated results of  magneto-energy spectrum, the ST transitions and geometry-engineered spin phase diagrams of two-electron charged quantum dots embedded in nanowires. Concluding remarks are presented in Sec. IV.

\section{Model}

\subsection{Single-particle model}

We begin with the problem of a single electron in a NWQD with a uniform
longitudinal magnetic field ${\bf B}=(0,0,B)$, which is described by the single-electron Hamiltonian,
\begin{equation}
H_0 =\frac{1}{2m^\ast}(\vec{p}+e\vec{A})^2 + V_c(x,y,z) + H_{\rm Z}.
\label{H_sp}
\end{equation}
Here the first term indicates the term of kinetic energy with ${\vec A}=(B/2)(-y,x,0)$ being the vector potential in symmetric gauge,
$m^\ast$ the effective mass of electron and $e$ the charge of an
elctron. The second term is the confining potential of NWQD modeled by
\begin{equation}
V_c(x,y,z)=\frac{1}{2}m^\ast \left[ \omega_0^2 \left( x^2 + y^2
\right) +\omega_z^2 z^2 \right]
\end{equation}
 with $\omega_0$ and $\omega_z$ parametrizing, respectively, the transverse
and the longitudinal confining strength. The last term is the
spin Zeeman energy $H_{\rm Z} =  g^\ast \mu_{B}B s_z$, in terms of
the $z$-component of electron spin $s_z=\pm1/2$, the effective Lande
$g$-factor of electron $g^\ast$ and the Bohr magneton $\mu_B$.
\begin{figure}[tp]
\includegraphics[width=0.45\textwidth,angle=0]{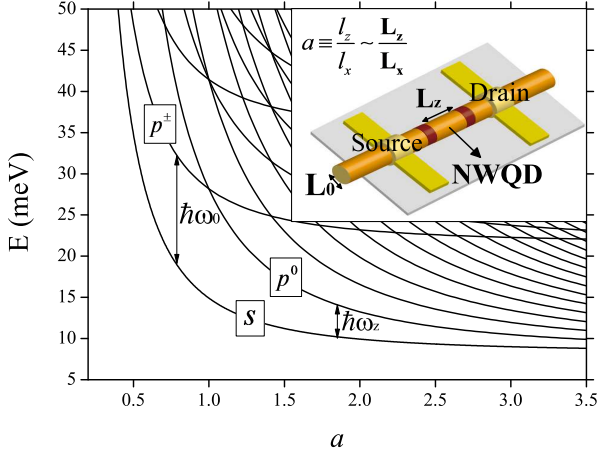}
\caption{(Color online) Single-electron energy spectrum as a function of aspect ratio $a$ of a NWQD with fixed lateral confinement $\hbar\omega_0=13.3$~meV at zero magnetic field obtained from the 3D parabolic model.
The considered lateral confinement strength $\hbar\omega_0=13.3$~meV corresponds to the cross section diameter ${\bf L}_0\sim 65$~nm for a cylindrical InAs nanowire.  The low-lying $s$-, $p^{\pm}$-, and $p^0$-orbitals are relevant to a two-electron problem.  The  energy quantization  for a short (long) NWQD with $a<1$ ( $a>1$) is characterized by the energy difference between the lowest and first excited orbitals $\hbar\omega_0$ ($\hbar\omega_z$).} \label{para_FD}
\end{figure}
The single-particle Hamiltonian (\ref{H_sp}) leads to the extended
Fock-Darwin single-particle spectrum
\begin{eqnarray}
\epsilon_{n,m,q,s_z}&=& \hbar\omega_{+}\left(n+\frac{1}{2}\right) +
\hbar\omega_{-}\left(m+\frac{1}{2}\right) \nonumber \\
&& + \hbar\omega_{z}\left(q+\frac{1}{2}\right) + E_{\rm Z}
\label{spspec}
\end{eqnarray}
where $n,m,q=0,1,2\cdots$ denote oscillator quantum numbers,  $E_{\rm Z}= g^\ast \mu_{B}B s_z$ is the
spin Zeeman energy, $\omega_{\pm}=\omega_h \pm \omega_c/2$ are in terms of the hybridized frequency $\omega_h\equiv (\omega_0^2+\omega_c^2/4)^{1/2}$ and the cyclotron frequency $\omega_c={eB}/{m^\ast}$. The corresponding eigenstate
$|n,m,q\rangle$ possesses the orbital angular momentum projection ${\ell}_z=\hbar
(n-m)$ and the parity $P=1$ ($P=-1$) with respect to $z$-axis for an
even (odd) $q$ number. The wave function of the lowest orbital is
given by
\begin{eqnarray}
 \psi_{000}(\vec{r})&=&  \left[(2 \pi)^{3/4} {l_h} \sqrt{l_z}\right]^{-1}\nonumber \\
  && \times {\rm{exp}} \left[-\frac{1}{4} \left( \frac{x^2+y^2}{l_h^2} + \frac{z^2}{l_z^2}\right)\right]
 \, ,
\label{psi000}
\end{eqnarray}
with the characteristic lengths of the wave function extents
$l_h=\sqrt{{\hbar}/{2 m^\ast \omega_h}}$ and $l_z=\sqrt{{\hbar}/{2
m^\ast \omega_z}}$.  The wave functions of other excited states can be
generated by successively applying the following defined raising
operators \cite{Hawrylak03}
\begin{eqnarray}
a^\dag &=& \frac{1}{\sqrt{2}} \left[\frac{x+iy}{2l_h}-l_h(\partial_x + i\partial_y)\right], \nonumber \\
b^\dag &=& \frac{1}{\sqrt{2}} \left[\frac{x-iy}{2l_h}-l_h(\partial_x + i\partial_y)\right],\\
a_z^\dag &=& \frac{z}{2l_z}-l_z\partial_z \nonumber
\end{eqnarray}
onto the ground state $|0,0,0\rangle$, i.e.
\begin{equation}
 |n,m,q\rangle =\frac{(\hat{a}^\dag)^n (\hat{b}^\dag)^m (\hat{a_z}^\dag)^q}{\sqrt{n!m!q!}}|0,0,0\rangle.
\end{equation}

The diameter of cross section of bottom-up synthesized nanowire is
typically $\sim 50-70$~nm. By contrast, the length of a QD in a
nanowire, defined by imposed electrodes or heterostructure potential
barriers, is highly tunable over a wide range from $10$ to
$300$~nm.~\cite{Bjork05}  For characterizing the geometry of a NWQD,
it is convenient to define the parameter of aspect ratio,
\begin{equation}
a \equiv \frac{l_z}{l_0}=\sqrt{\frac{\omega_0}{\omega_z}}
\label{def_a}
\end{equation}
according to the characteristic length of the lowest orbital wave
function based on the 3D parabolic model.  A rod-like
(disk-like) NWQD is characterized by the value of aspect
ratio $a>1$ ($a<1$), where the longitudinal extent of the
electron wave function is longer (shorter) than the transverse one on the cross
section of the nanowire. Notably, the effective aspect ratio $a = l_z/l_0$ defined here is not but very close to the geometric aspect ratio $a_{\rm
geom}$, namely $a \simeq a_{\rm geom}= {\bf L}_z/{\bf L}_0$ with ${\bf L}_0$ (${\bf L}_{z}$) being the cross section diameter (length) of NWQD.
\begin{widetext}
\begin{figure*}
\includegraphics[width=0.9\textwidth,angle=0]{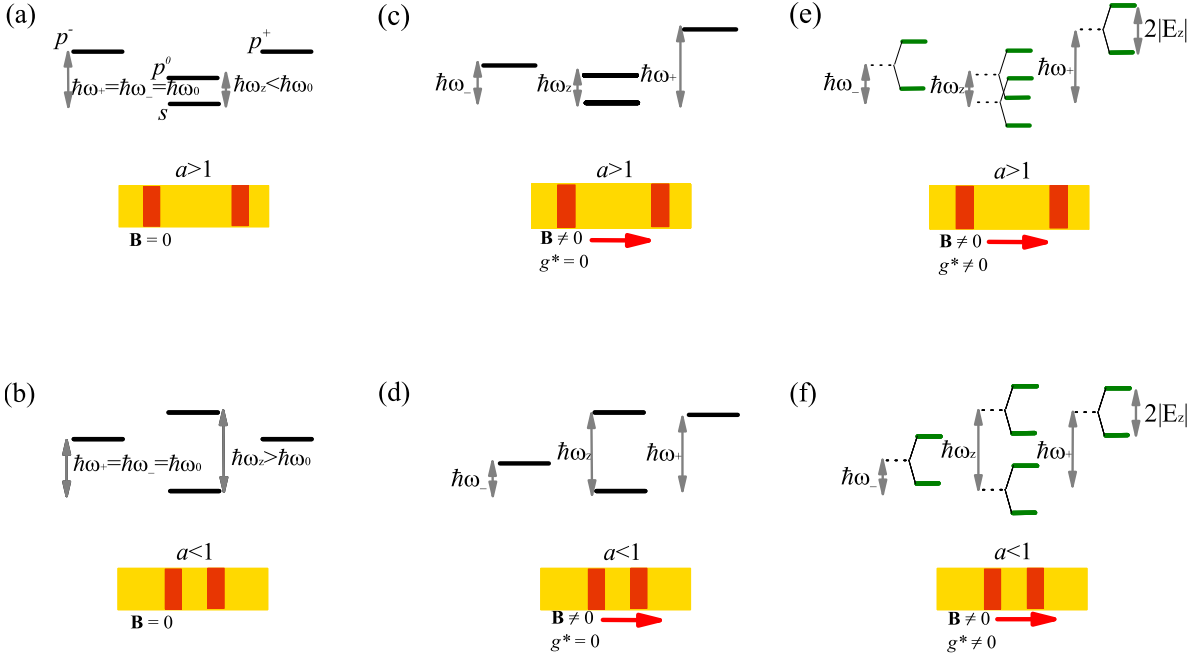}
\caption{(Color online) Schematic illustration of the electronic structures,
consisting of few relevant low lying orbitals (one $s$- and three $p$-orbitals), of long rod-like NWQDs [(a)(c)(e)] and short disk-like NWQDs
[(b)(d)(f)] with or without longitudinal magnetic field $B$ and including
or excluding the spin Zeeman splitting $E_{\rm{Z}}$ ($g^\ast =0$ or
$g^\ast \neq 0$). (a) $a>1$ and $B=0$; (b) $a<1$ and $B=0$; (c) $a>1$, $B\ne0$ and $g^ \ast
=0$; (d)$a<1$, $B\ne0$ and $g^ \ast =0$; (e)$a>1$, $B\ne0$ and $g^ \ast \ne 0$; (f)$a<1$, $B\ne0$ and $g^ \ast \ne
0$.} \label{sys_illus}
\end{figure*}
\end{widetext}
Figure~\ref{para_FD} presents the calculated single-electron energy spectrum as a function of aspect ratio $a$ for a NWQD with fixed lateral confinement
$\hbar\omega_0=13.3$~meV at zero magnetic field according to Eq.(\ref{spspec}). The chosen parameter of lateral confinement $\hbar\omega_0=13.3$~meV is determined
by fitting the numerically calculated energy separation between the two lowest single-electron orbitals of a cylindrical InAs/InP NWQD of cross section diameter
${\bf L}_0=65$~nm by 3D finite difference simulation. In the simulation, the $\rm{Schr\ddot{o}dinger}$ equation for a single electron confined in a 3D cylindrical potential
well is solved by using finite difference method, with the used parameters: the effective mass $m^\ast=0.023m_0$ of electron for InAs and the InAs/InP band edge offset
$V_b=0.6$~eV as the barrier height of the confining potential.~\cite{Bjork05,Bjork02}

In a two-electron (2e) problem, the most relevant orbitals are the two lowest ones because the kinetic energy difference between the two orbitals
is the main energy cost, in competition with the coulomb or
spin Zeeman energies, for a spin triplet state to be the ground state of two-electron.
By convention, we from now on name the lowest single electron state $|n,m,q\rangle = |0,0,0\rangle$ as $s$-orbital, and the next three $p$-shell states $|0,0,1\rangle$, $|1,0,0\rangle$, and  $|0,1,0\rangle$ as $p^0$-, $p^+$-, and $p^-$-orbitals, respectively. According to \eq{spspec}, the energy of the lowest $s$-orbital is explicitly given by
\begin{equation}
\epsilon_{s,s_z}
=\frac{1}{2}\left(\hbar \omega_{+}+\hbar \omega_{-}+\hbar \omega_{z}\right)+g^\ast \mu_{B}B s_z,
\end{equation}
and those of the three p-shell orbitals are respectively given by
\begin{eqnarray}
&& \epsilon_{p^0,s_z}
=\epsilon_{s,s_z}+\hbar\omega_z, \nonumber \\
&&\epsilon_{p^+,s_z}
= \epsilon_{s,s_z}+\hbar\omega_{+}, \nonumber \\
&&\epsilon_{p^-,s_z}
= \epsilon_{s,s_z}+\hbar\omega_{-}.
\label{e_s_ps}
\end{eqnarray}

For $B=0$, we have $\epsilon_{s,s_z}=\hbar \omega_{0}\left(1+1/2a^2\right)$, $\epsilon_{p^0,s_z}=\epsilon_{s,s_z}+ \hbar \omega_{0}/a^2$, and $\epsilon_{p^+,s_z}=\epsilon_{p^-,s_z}= \epsilon_{s,s_z}+ \hbar \omega_{0}$ according to Eqs.(\ref{def_a}) and (\ref{e_s_ps}). Here, the $p^+$- and $p^-$-orbitals are degenerate with the same energy separation from the $s$-orbital, $\hbar\omega_{\pm}=\hbar\omega_0$, while the $p^0$-orbital is energetically higher than $s$-orbital by $\hbar\omega_z=\hbar\omega_0/a^2$. Obviously, $p^0$ ($p^{\pm}$) is the second lowest orbital for a long (short) NWQD with $a>1$ ($a<1$) at zero magnetic field, as shown in Fig.~\ref{para_FD}. For a symmetric NWQD with $a=1$, the $p^0$-  and $p^{\pm}$-orbitals form a 3-fold orbital-degenerate shell. Figure~\ref{sys_illus} (a) [(b)] schematically depicts the low-lying orbitals of a long [short] NWQD with  $a>1$ [$a<1$] at zero magnetic field.

Applying a longitudinal magnetic field onto a cylindrical NWQD breaks the degeneracy of $p^+$- and $p^-$-orbitals.
The orbital Zeeman effect lowers (raises) the energy level of the $p^-$($p^+$)-orbital from $\hbar\omega_{0}$ to $\hbar\omega_{-}$ ($\hbar\omega_{+}$). Thus, if a long NWQD is subjected to a sufficiently strong magnetic field, the second lowest orbital of the dot could be changed from the $p^0$ to $p^-$. By contrast, the  second lowest orbital of a short NWQD is always the $p^-$-orbital.
Therefore, the characteristic energy quantization of the $p^-$-orbitals,
$\hbar \omega_{-} = \hbar\left( \omega_0^2 + \omega_c^2/4
\right)^{1/2} - \hbar\omega_c/2$, is often a key parameter for a short NWQD or a moderately long NWQD with strong magnetic field.
Considering wide-band gap materials such as GaAs, the $g$-factors are usually small and the spin Zeeman effect on the energy shift of orbital is negligible.
Figure~\ref{sys_illus}(c) [(d)] depicts the $B$-dependent electronic orbitals of a long [short] NWQDs, where vanishing spin Zeeman
splitting is assumed ($g^\ast=0$ is set).

For a low energy gap material with larger $g^\ast$, like InAs, the spin Zeeman effect could be significant in the spin ST  transition of two-electron QD.
Figure~\ref{sys_illus}(e) [(f)] schematically shows the spin-resolved electronic orbitals of a long [short] NWQDs with $B\neq 0$ and $g^\ast\neq 0$ by the spin Zeeman
splitting $2E_{\rm Z}$. With the spin Zeeman effect, all the spin-up (spin-down) orbitals are energetically lowered (raised) by $E_{\rm Z}=g^\ast \mu_B B/2$ according to
\eq{spspec}. If the applied magnetic field or the $g$-factor of material is so large that the spin Zeeman splittings exceed the kinetic energy quantization of QD,
both of the two lowest single-electron states are the spin-up ones and the ground state of the 2e dot is ensured to be a spin triplet state simply according to
spin Pauli exclusion principle.

In this work, the following formulation for the $g$-factor of an InAs-based QD is adopted~\cite{Hermann77,Bjork05}
\begin{equation}
 g^\ast = g \left[
        1 - \frac{P^2}{3}\frac{\Delta_{\rm SO}}{E_g^{\rm eff}\left(E_g^{\rm eff}
        + \Delta_{\rm SO}\right)}
        \right]\, , \label{geff}
\end{equation}
where $E_g^{\rm eff}$ is the effective energy gap of semiconductor
QD, $g=2.0$ is the Lande $g$-factor for free electron, $\Delta_{\rm
SO}$ is the spin-orbit splitting in the valence band, and $P$ is the parameter of interband transition matrix element.~\cite{Hermann77}

Here, the effective energy gap of a QD can be estimated as $E_g^{\rm eff} =
E_g^{\rm bulk}+\epsilon_{s,s_z}$, where  $E_g^{\rm bulk}$ is the bulk energy
gap and $\epsilon_{s,s_z}$ is the quantization energy of the lowest electronic orbital of the QD with $B=0$ measured from the conduction band edge. For InAs-based QDs, we take the following parameter values: $E_g^{\rm bulk}=460$~meV, $\Delta_{\rm SO}=390$~meV,
$P^2=21.5$~eV.~\cite{Bjork05} Accordingly, the value of $g^\ast$ for a symmetric NWQD with ${\bf L}_0={\bf L}_z=65$~nm is estimated as large as $g^\ast \approx -11$.~\cite{Bjork05}

\subsection{Interacting few-electron model}

The interacting Hamiltonian of few electrons in a NWQD can be
expressed in the form of second quantization as
\begin{eqnarray}
H &=& \sum_{i,\sigma}\epsilon_{i\sigma}
c_{i\sigma}^{\dag}c_{i\sigma}\nonumber \\
&&+ \frac{1}{2}\sum_{ijkl,\sigma \sigma'}\langle ij|V|kl\rangle
c_{i\sigma}^{\dag}c_{j\sigma'}^{\dag}c_{k\sigma'}c_{l\sigma} \, ,
\label{full_H}
\end{eqnarray}
where $i,j,k,l$ denote the composite indices of single electron
orbitals such as $|i\rangle=|n_i,m_i,q_i\rangle$, $c_{i\sigma}^{\dag}$ ($c_{i\sigma}$) the electron creation
(annihilation) operators, and $\sigma=\pm$ the electron
spins $s_z=\pm \frac{1}{2}$.  The first (second) term on the right
hand side of Eq.(\ref{full_H}) represents the kinetic energy of
electrons (the Coulomb interactions between electrons) and the
Coulomb matrix elements are defined as
\begin{eqnarray}
\langle ij|V|kl\rangle &\equiv& \frac{e^2}{4\pi\kappa}  \int \int
d\vec{r_1} d\vec{r_2}
\psi_i^*(\vec{r_1})\psi_j^*(\vec{r_2}) \nonumber \\
&&\times \frac{1}{|\vec{r_1} - \vec{r_2}|}\psi_k(\vec{r_2})
\psi_l(\vec{r_1}) \, ,
\end{eqnarray}
where $\kappa $ is the dielectric constant of dot material. For InAs material, we take $\kappa=15.15 $.
After lengthy derivation, one can obtain the generalized Coulomb matrix elements for the case of $a\ge 1$:
\begin{widetext}
\begin{eqnarray}
& &\langle
n_{i}m_{i}q_{i};n_{j}m_{j}q_{j}|V|n_{k}m_{k}q_{k};n_{l}m_{l}q_{l}\rangle
= \nonumber
\\
& &\left(\frac{1}{\pi l_h}\right)    \frac{
\delta_{R_L,R_R} \cdot \delta_{q_i+q_j+q_l+q_k, \rm{even}}
  }{\sqrt{n_i!m_i!q_i!n_j!m_j!q_j!n_k!m_k!q_k!n_l!m_l!q_l!}} \nonumber\\
& &\times \sum_{p_1=0}^{\min(n_i,n_l)} \sum_{p_2=0}^{\min(m_i,m_l)}
\sum_{p_3=0}^{\min(q_i,q_l)} \sum_{p_4=0}^{\min(n_j,n_k)}
\sum_{p_5=0}^{\min(m_j,m_k)} \sum_{p_6=0}^{\min(q_j,q_k)} \
p_1!p_2!p_3!p_4!p_5!p_6!\nonumber \\
& &\times
{n_i \choose p_1}
{n_l \choose p_1}
{m_i \choose p_2}
{m_l \choose p_2}
{q_i \choose p_3}
{q_l \choose p_3}
{n_j \choose p_4}
{n_k \choose p_4}
{m_j \choose p_5}
{m_k \choose p_5}
{q_j \choose p_6}
{q_k \choose p_6}\nonumber\\
& &\times (-1)^{u+v/2+n_j+m_j+q_j+n_k+m_k+q_k}  {\left(\frac{1}{2}\right)^u}  x^{u+1/2}\nonumber\\
& &\times
\frac{\Gamma(\frac{1+2u+v}{2})\Gamma(1+u)\Gamma(\frac{1+v}{2})}{\Gamma(\frac{3+2u+v}{2})}
 {_2F_1}\left(1+u,\frac{1+2u+v}{2};\frac{3+2u+v}{2};1-x\right)\,  ,
\label{Vijkl}
\end{eqnarray}
\end{widetext}
where we define $u=m_i+m_j+n_l+n_k-(p_1+p_2+p_4+p_5)$,
  $v=(q_i+q_l+q_j+q_k)-2(p_3+p_6)$,
  $R_L=(m_i+m_j)-(n_i+n_j)=-(\ell_{z,i}+\ell_{z,j})$,
  $R_R=(m_l+m_k)-(n_l+n_k)=-(\ell_{z,l}+\ell_{z,k})$,
$x\equiv \omega_z/\omega_h$, and $_2F_1$ is the hypergeometric function. The $\delta$-functions
$\delta_{q_i+q_j+q_l+q_k, \rm{even}}$ and $\delta_{R_L,R_R}$ in the
formulation ensure the conservation of the parity with respect to
$z$-axis and the $z$-component of angular momentum of system $L_z$,
respectively. The formulation of \eq{Vijkl} is confirmed by computing the Coulomb integral numerically.

For short NWQDs with  $a<1$, the formulations of the Coulomb matrix
elements are obtained by simply taking Euler's hypergeometric
transformation for the hypergeometric function in \eq{Vijkl}, i.e.,
replacing
\begin{equation}
{_2F_1}\left(1+u,\frac{1+2u+v}{2};\frac{3+2u+v}{2};1-x\right)\, \nonumber
\end{equation}
by
\begin{equation}
 x^{-\frac{1+2u+v}{2}} {_2F_1}\left(\frac{1+v}{2},\frac{1+2u+v}{2};\frac{3+2u+v}{2};1-\frac{1}{x}\right)\,. \nonumber
\end{equation}
The generalized formulations for the Coulomb matrix elements based
on the 3D asymmetric parabolic model are probably for the first time
derived, which allows for straightforward implementation of the CI
theory and is widely applicable to arbitrary 3D confining
semiconductor nanostructures.

\subsection{Exact diagonalization}
Based on the CI theory presented above, we follow the standard numerical exact diagonalization procedure to calculate the energy spectrum of $N_e$ interacting electrons in a NWQD .~\cite{Hawrylak03}
The numerically exact results are obtained
by increasing the numbers of chosen single electron orbital basis and the corresponding $N_e$-electron configurations until a numerical convergence is achieved.
In the full configuration interaction (FCI)
calculation for a 2e problem, we usually take the number of single electron orbitals typically from $20$ to $26$ and that of the corresponding 2e configurations
from $190$ to $325$ to have a satisfactory numerical convergence.

\section{Numerical results and discussion}

\subsection{Magnetic-field induced ST transitions}
\begin{figure}
\includegraphics[width=0.45\textwidth,angle=0]{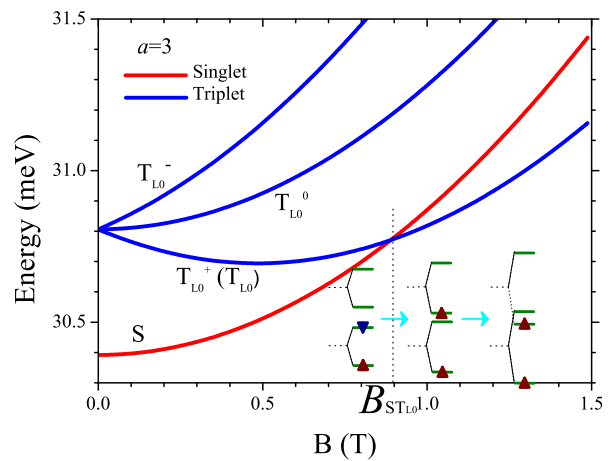}
\caption{(Color online) Magneto-energy spectrum of two interacting electrons in a
NWQD with transverse confining strength $\hbar \omega_0=13.3$~meV and aspect ratio $a=3$.} \label{2eEB}
\end{figure}

Let us first consider two interacting electrons in a rod-like NWQD with the aspect
ratio $a=3$ and the transverse confining strength $\hbar
\omega_0=13.3$~meV using FCI calculation. The low-lying magneto-energy
spectrum of the two-electron NWQD is shown in \fig{2eEB}, which consists of a spin singlet state branch, labeled by $\rm{S}$, and three triplet state branches split by the spin Zeeman energy, labeled by $\rm{T_{L0}^+}$, $\rm{T_{L0}^0}$ and $\rm{T_{L0}^-}$ according to the z-component of total spin ($S_z=+1$, $S_z=0$ and $S_z=-1$), respectively.~\cite{Fasth07,Hanson07}
Since usually only triplet states with $S_z=+1$ are involved in ST transitions, we shall use $\rm T_{\rm L \it {\left|L_z\right|}}$ to denote the triplet states
with angular momentum $L_z$ through out this article, skipping the
superscript $+$ of $\rm{T_{L \it{\left|L_z\right|}}^+}$ for brevity.

The main configurations of the two-electron
ground states around the critical magnetic field are schematically
shown in the lower right corner of \fig{2eEB}. In the weak magnetic field regime $\left(B<B_{\rm{ST_{L0}}} \sim 0.9~\rm T\right)$, the two electrons in the NWQD
mainly occupy the lowest $s$-orbital simply following the Aufbau principle,
and form a spin singlet ground state. With increasing
$B$, the triplet state $\rm{{T_{L0}}}$ is more energetically
favorable than the singlet state because of the increasing spin
Zeeman energy, the reduced Coulomb repulsion, and exchange energy between the two spin polarized
electrons. A crossing of the singlet branch and the triplet state
branch $\rm{T_{L0}}$ is observed at the critical magnetic field
$B_{\rm{ST_{L0}}}=0.9$~T. Such magnetic-field induced ST transitions are attributed to
the energetic competition between single particle energy
quantization, the spin Zeeman energy, and the various Coulomb
interactions including the direct, exchange, and correlation
interactions as well.~\cite{Hawrylak93-prl}

Other weak spin-related terms, such as the spin-orbital coupling (SOC) with 1-2 order of magnitude smaller than the kinetic quantization of QD are neglected in the Hamiltonian
of Eq.(\ref{full_H}). The SOC mixes the spin of the $\rm{S}$ and $\rm{{T_{L0}}}$ states and creates an anti-crossing of the
S- and $\rm{{T_{L0}}}$-branches around the $B_{ST}$ with a small energy gap, typically only $\sim 0.1-0.5$~meV as observed in previous experiments.~\cite{Fasth07}

\subsection{Spin phase diagram}

\begin{figure}
\includegraphics[width=0.45\textwidth,angle=0]{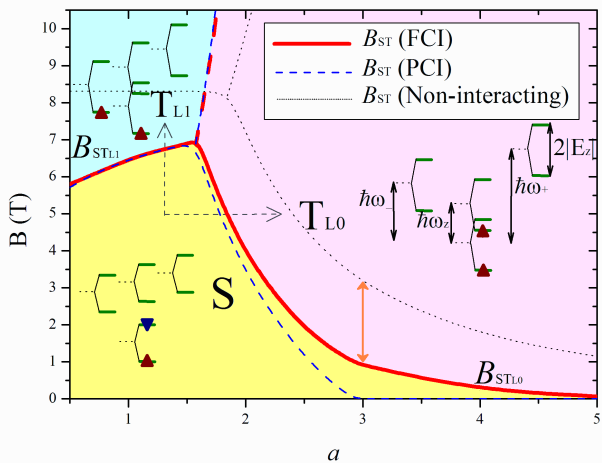}
\caption{(Color online) Spin phase diagrams of two-electron NWQDs of lateral
confinement $\hbar \omega_0=13.3$~meV with respect to tunable magnetic
field $B$ and aspect ratio $a$. The phases are distinguished by the
curves of critical magnetic field $B_{\rm{ST}}$ obtained from non-interacting (black dotted), PCI (blue dashed), and FCI (red solid) calculations.}
\label{a_B_phase_extended}
\end{figure}

Figure~\ref{a_B_phase_extended} shows the calculated spin phase diagrams of the
two-electron ground state of the NWDQs with a fixed cross section diameter (fixed $\hbar \omega_0=13.3$~meV) but various lengths (various $\hbar\omega_z$) with respect to the applied magnetic field $B$ and
the aspect ratio $a$. Three phases ($\rm S$, $\rm{T_{L0}}$, and $\rm{T_{L1}}$) are distinguished by the
curves of critical magnetic field $B_{\rm{ST}}$ in Fig.~\ref{a_B_phase_extended}. Correspondingly, the main configurations of the 2e ground states are depicted inside the
colored regions of the phases. To identify the various underlying mechanisms in the phase diagrams, including the spin Zeeman effect and the inter-particle Coulomb interactions,
the spin phase diagrams are calculated by using non-interacting, full CI, and partial CI calculations, respectively.

In the non-interacting calculation,
the coulomb interactions are artificially disabled and the considered ST
transitions are induced only by the spin Zeeman effect. The
comparison between the results of non-interacting and FCI calculations allows us to distinguish effects of the Coulomb interaction and spin Zeeman
coupling on the ST transitions. In particular, to highlight the
Coulomb correlation effect, a partial configuration interaction (PCI) calculation is also performed for the
spin phase diagrams, in which only the lowest energy configuration is taken as the sole basis and the couplings from higher
energy configurations are excluded.

The essential features of the phase diagrams can be realized based on
the non-interacting picture. For a not very long (small or moderate $a$) NWQD with weak $B$, the 2e ground state is likely to be the spin singlet state $\rm S$, simply following Aufbau principle (the yellow region in \fig{a_B_phase_extended}).
Starting from the singlet phase $\rm S$, the two-electron ground state of a NWQD might be switched to the spin triplet phases (the pink region $\rm{T_{L0}}$ or the cyan region $\rm{T_{L1}}$) by
increasing either $a$ or $B$ (see the horizontal and vertical dashed lines with arrows in \fig{a_B_phase_extended} for the guidance of eyes).

Following the horizontal dashed line, the longitudinal energy quantization $\hbar \omega_z$ is decreased by the increase of $a$.
With the addition of spin Zeeman term, the spin-up level of $p^0$-orbital could become even lower than the spin-down level of $s$-orbital if the decreasing $\hbar \omega_z$
is so small as that $\hbar \omega_z<2|E_z|$ (see the difference between the schematic configurations for the $\rm S$ and $\rm{T_{L0}}$ states).
In this situation, the 2e ground state can transit to the spin triple states $\rm{T_{L0}}$, simply following spin Pauli exclusion principle.
On the other hand, the transition of a 2e ground state of NWQD from the singlet state $\rm S$ to the triple one $\rm{T_{L1}}$ is shown also possible by increasing the strength
of applied magnetic field. Following the vertical  dashed line, increasing $B$ reduce the energy separation between the $s$- and $p^-$-orbital levels, i.e. $\hbar \omega_{-}$. Similar to the case of $\rm{S}$-$\rm{T_{L0}}$ transition, a $\rm S$-$\rm{T_{L1}}$ transition can happen as the decreased $\hbar \omega_{-}$ is so small as that $\hbar \omega_{-}<2|E_z|$ .
 In the non-interacting picture, $B_{\rm ST_{L0}}$ is explicitly given by $B_{\rm ST_{L0}}=\hbar\omega_0/g^ \ast \mu_B a^2$, showing a quadratic decay with $a$, while the critical magnetic field $B_{\rm ST_{L1}}$ for $\rm{S}$-$\rm{T_{L1}}$ transitions is
dependent only on $\hbar \omega_0$ and remains nearly constant in the $B-a$ plot.

The Coulomb interactions are shown to reduce the singlet phase area in
the diagrams from the comparison between the non-interacting and CI results.
For example, the segment of vertical solid line at $a=3$ in \fig{a_B_phase_extended} indicates that the critical magnetic field is significantly reduced
from $B_{\rm ST_{L0}}(\rm {Non-interacting})=3.2$~T to $B_{\rm ST_{L0}}(\rm{FCI})=0.9$~T as the Coulomb interactions are taken into account.
This is because the spin triplet states gain additional negative
exchange energies while the singlet state does not. We also notice that
the $B_{\rm ST_{L1}}$ for the $\rm{S}$-$\rm{T_{L1}}$ transition no
longer remains constant but slightly increases with increasing $a$ because the
strength of the coulomb interactions is reduced by the increase of dot
volume.

Basically, the results obtained from the FCI and PCI calculations
have similar features except for those in the regime of high $a$ ($a>3$). While the PCI calculation
shows the vanishing $B_{\rm ST_{L0}}$ for $a\sim 3$, the FCI
calculation yields the always non-zero $B_{\rm ST_{L0}}$. This means
that the Coulomb correlations energetically favor the spin singlet
state as ground state and become more pronounced in long NWQDs.

\subsection{Crossover from disk-like to rod-like QDs}

The spin phase diagrams of Fig.~\ref{a_B_phase_extended} suggest that
purposely accessing a specific spin phase of two-electron is
feasible through the geometrical control of NWQDs. For instance, the ground state
of a two-electron NWQD can be switched from the singlet $\rm{S}$ to the triplet
state $\rm{T_{L0}}$ by increasing the aspect ratio $a$ at the fixed $B=5$~T (trace the horizontal dashed line in \fig{a_B_phase_extended}).

\begin{figure}
\includegraphics[width=0.5\textwidth,angle=0]{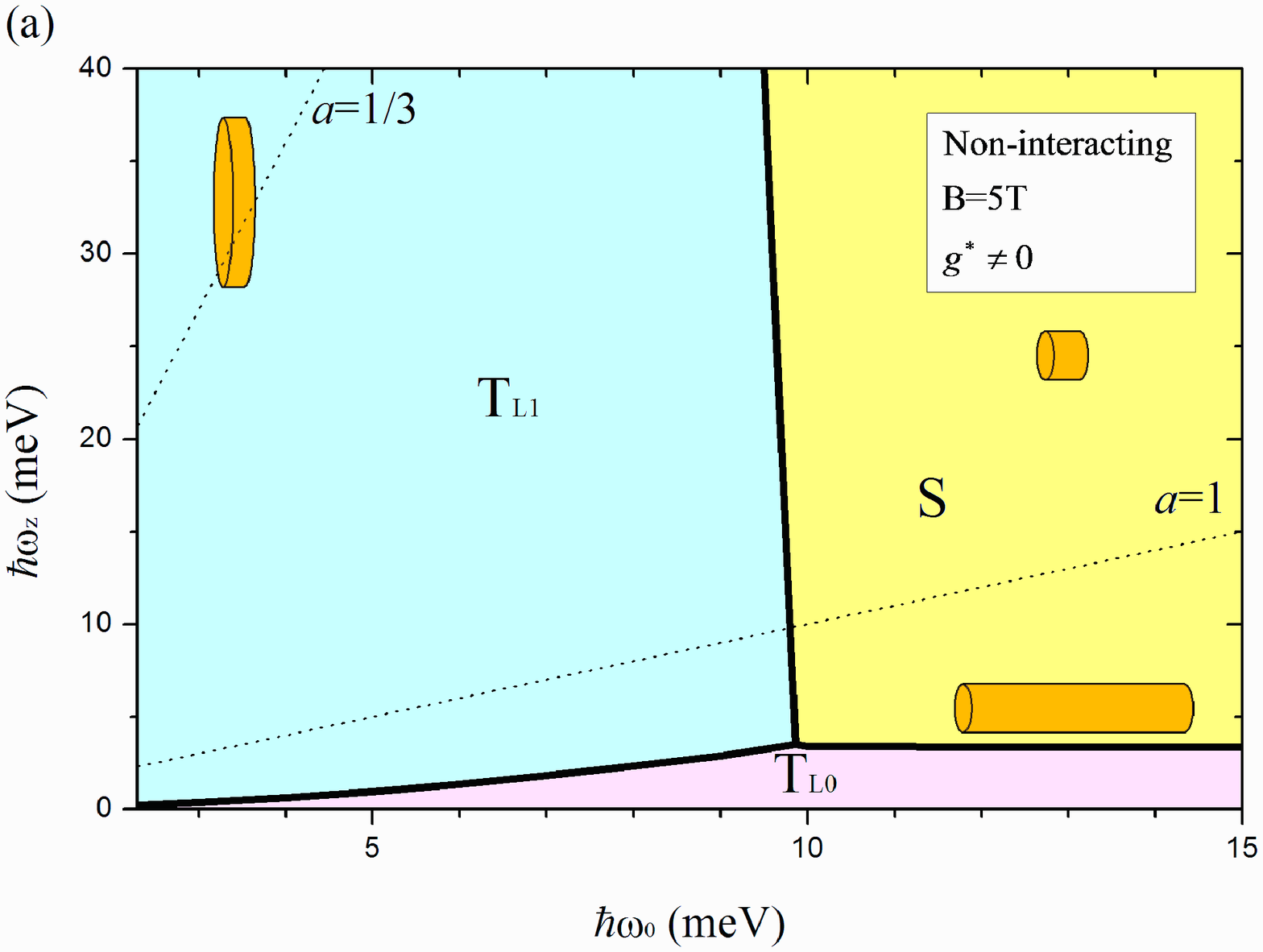}
\includegraphics[width=0.5\textwidth,angle=0]{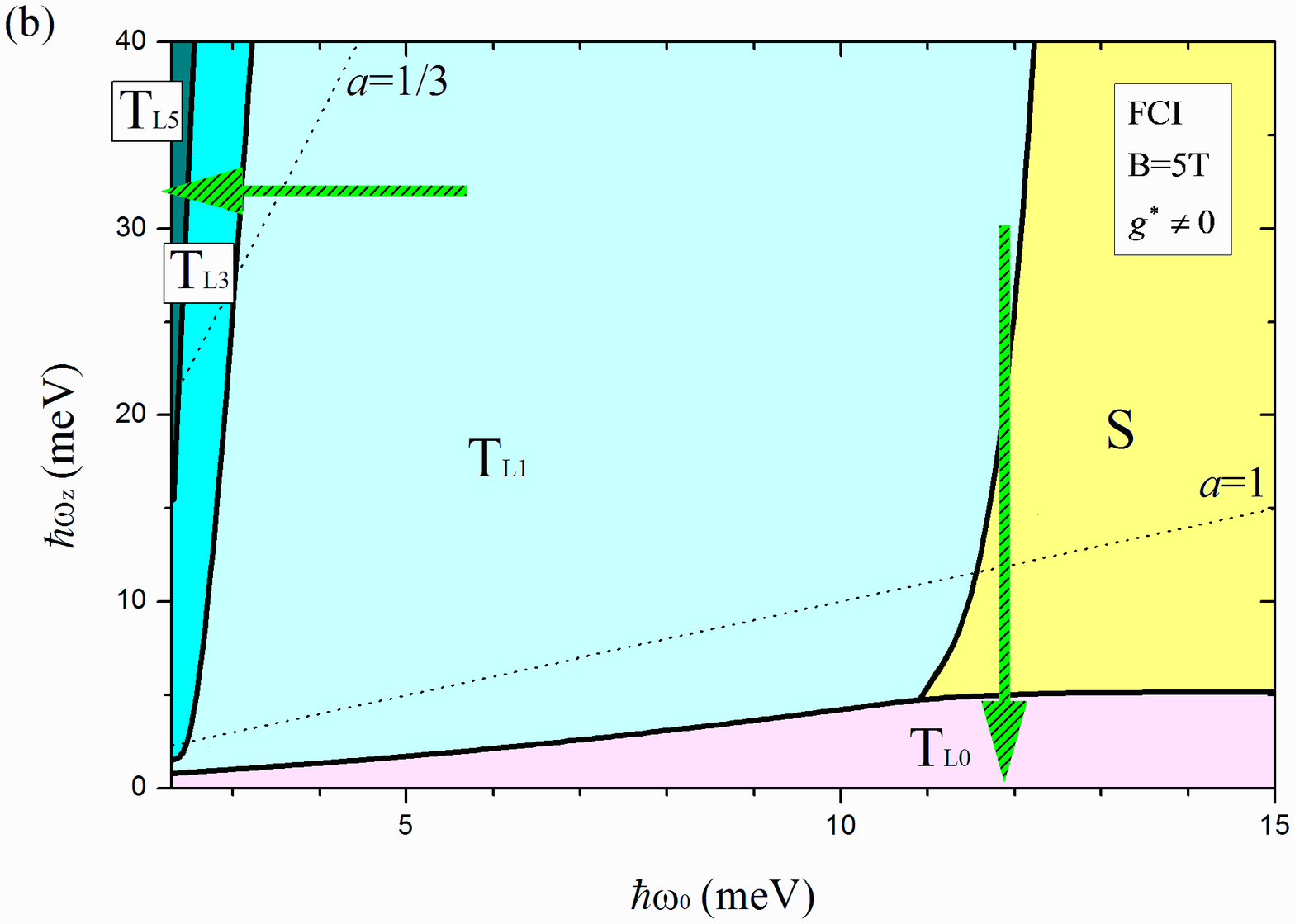}
\includegraphics[width=0.507\textwidth,angle=0]{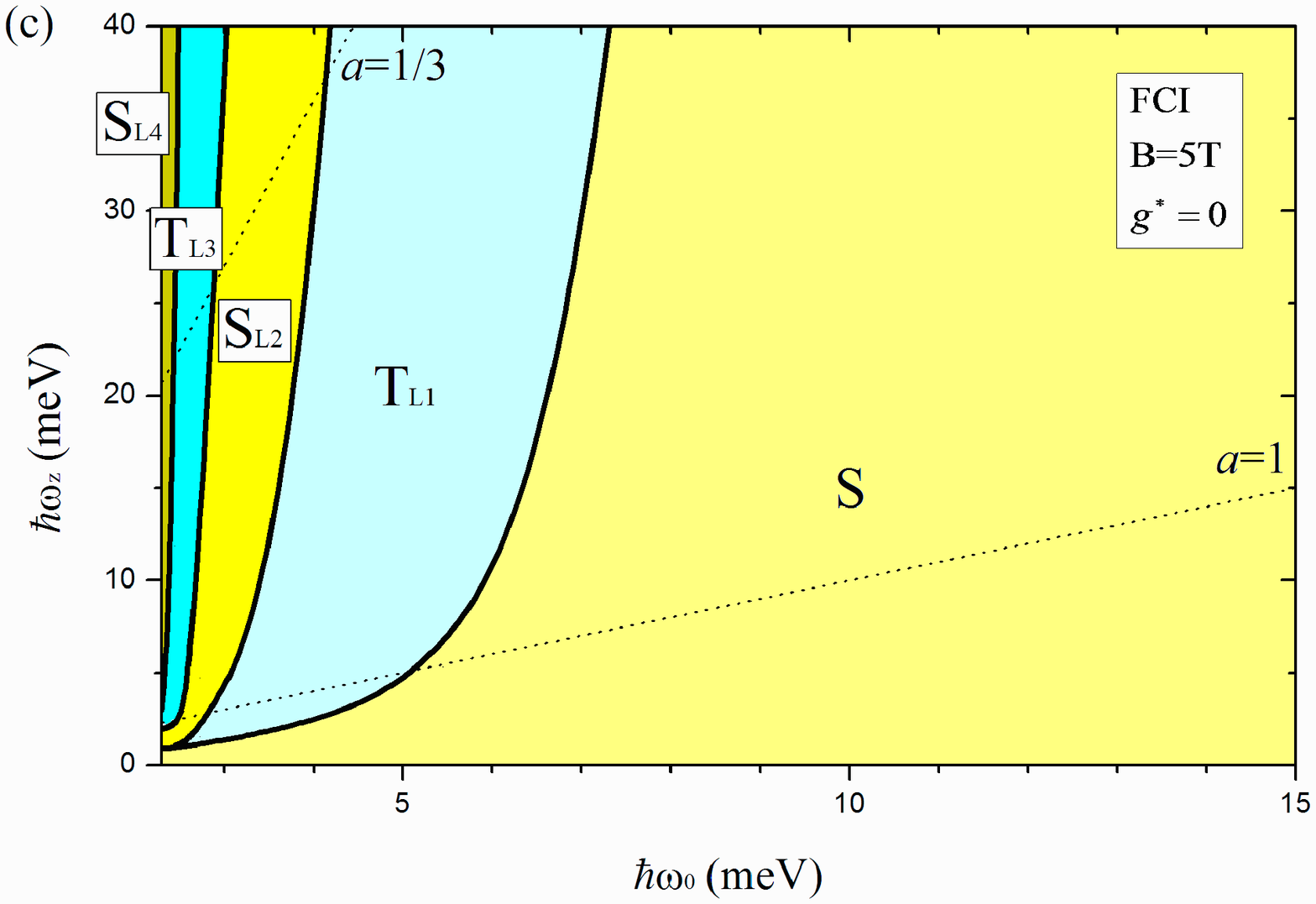}
\caption{(Color online) Spin phase diagrams of doubly charged NWQDs with respect to the
lateral and longitudinal confinements, parametrized by $\hbar \omega_0$
and $\hbar \omega_z$, respectively, in a fixed magnetic field $B=5$~T
for (a) non-interacting two electrons with $g^\ast \ne 0$, (b)
interacting two electrons with $g^\ast \ne 0$ and (c) interacting two
electrons with $g^\ast = 0$.}
\label{phase}
\end{figure}
\begin{figure}
\includegraphics[width=0.45\textwidth,angle=0]{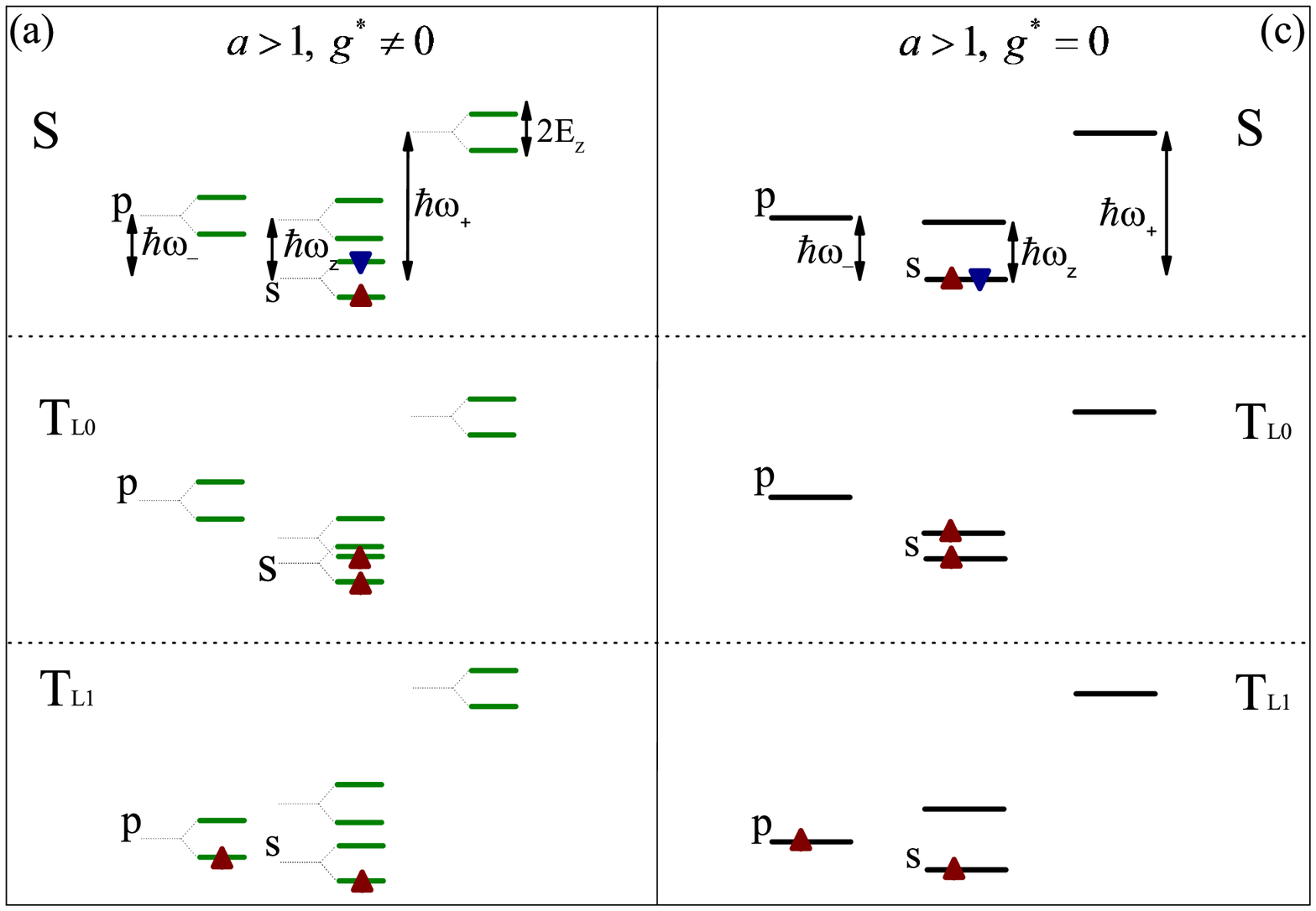}
\includegraphics[width=0.45\textwidth,angle=0]{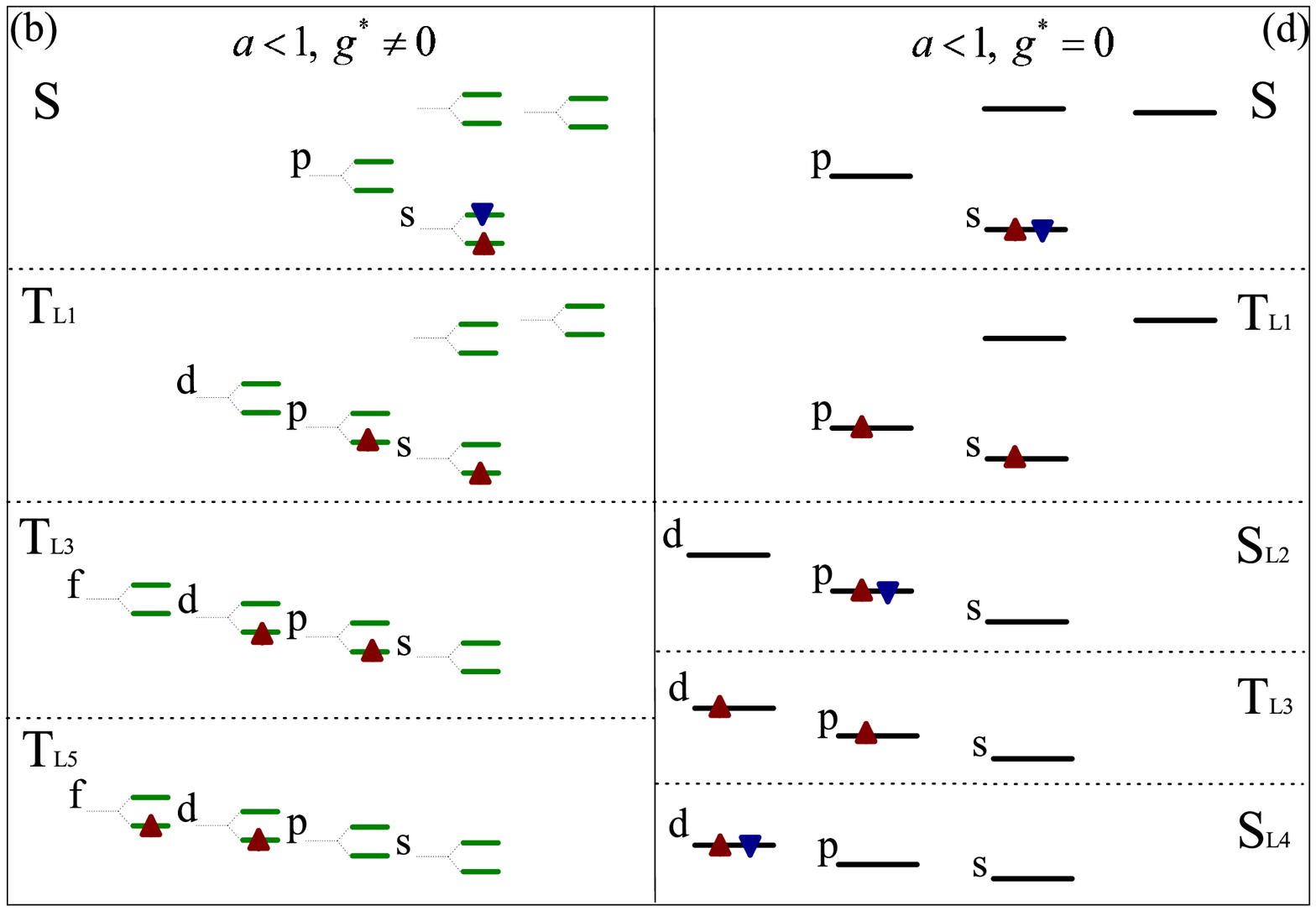}
\caption{(Color online) Relevant two-electron configurations possibly being the main
components in the ground states of NWQDs in an uniform magnetic field $B=5$~T for (a) $a>1$ and $g^\ast \ne 0$, (b) $a<1$ and
$g^\ast \ne 0$, (c) $a>1$ and $g^\ast =0$, and (d) $a<1$ and $g^\ast =0$.} \label{configs}
\end{figure}

Figure~\ref{phase} presents the spin phase diagrams of two-electron NWQDs with respect to the
lateral and longitudinal confinements, parametrized by $\hbar \omega_0$
and $\hbar \omega_z$, respectively, in a fixed magnetic field $B=5$~T
for (a) non-interacting two electrons with $g^\ast \ne 0$, (b)
interacting two electrons with $g^\ast \ne 0$, and (c) interacting two
electrons with $g^\ast = 0$.
In \figs{configs}(a) and (b), we present the relevant two-electron configurations to the spin phase diagrams of
\figs{phase}(a) and (b) with the inclusion of spin Zeeman effect $\left( g^\ast \ne 0 \right)$,
while in \figs{configs}(c) and (d) we present the relevant two-electron
configurations to the spin phase diagrams of \fig{phase}(c) for $g^\ast = 0$

The non-interacting spin phase
diagram is first shown in Fig.~\ref{phase}(a) in order to identify the spin Zeeman effect and also contrast the Coulomb interaction effects
on the interacting spin phase diagrams presented in \fig{phase}(b).
In the non-interacting case,
the features of the spin phases of Fig.~\ref{phase}(a) are purely
determined by the competition between geometry-dependent quantized
electronic structures of dots and the spin Zeeman splitting, which
is nearly a constant here created by the fixed $B$. Three distinctive spin phases, $\rm{S}$, $\rm{T_{L0}}$, and  $\rm{T_{L1}}$, are marked in different  colors in Fig.~\ref{phase}(a).
In the yellow region where both $\hbar \omega_0$ ($\hbar \omega_-$) and $\hbar \omega_z$ are large, the
kinetic quantizations in both longitudinal and lateral directions
are stronger than the spin Zeeman splitting and $\rm{S}$ remains as
a ground state. Reducing the longitudinal confinement, $\hbar \omega_z$, can lead to the $\rm{S}$-$\rm{T_{L0}}$ (from the yellow to the pink region) transition as $\hbar \omega_z \lesssim
2|E_{\rm{Z}}|$.
Similarly, reducing the transverse confinement leads to the $\rm{S}$-$\rm{T_{L1}}$ transition as $\hbar \omega_- \lesssim
2E_{\rm{Z}}$(from the yellow to the light cyan region).

Compared with
\fig{phase}(a), the interacting spin phase diagram of \fig{phase}(b)
shows the following additional features:

(i) Larger areas of both
$\rm{T_{L0}}$ and $\rm{T_{L1}}$ phases are observed because of the additional negative exchange energies and the reduced direct Coulomb repulsions gained by the triplet states.

(ii) A NWQD with $\hbar \omega_0\approx 12$~meV could  experience a three-phase transitions from $\rm{T_{L1}}$ (cyan) to $\rm{S}$ (yellow), and then to $\rm{T_{L0}}$ (pink) with increasing the length of wire, from $\hbar \omega_z>25$~meV to  $\hbar\omega_z<5$~meV (see the vertical line positioned at $\hbar \omega_0=12$~meV in \fig{phase}(b)).

(iii) In the regime of small $\hbar \omega_0$ and large
$\hbar \omega_z$ (i.e. flat quasi-2D dots with $a\ll 1$), a series of transitions from the spin single states to various triplet
states, $\rm{T_{L1}}$, $\rm{T_{L3}}$, $\rm{T_{L5}}$, etc. [see \fig{configs}(b)] and a staircase increase of total orbital angular momentum are observed with reducing the lateral confinement $\hbar\omega_0$.

In the weak laterally confining regime,
few electrons in the quasi-2D QD in a high magnetic field successively fill the orbitals with negative $z$-projection of
orbital angular momentum, i.e. the orbitals of lowest Landau
level (LLL), with small kinetic energy separation $\hbar\omega_{-}$.  The inter-particle Coulomb interactions thus become
particularly pronounced among the particles on the nearly
degenerate LLL orbitals with alomost quenched kinetic energies. In order to
minimize the coulomb repulsion, the particles on the quasi-degenerate
orbitals tend to spread the occupancy of orbitals as far as possible, but in
competition with the cost of increase of kinetic energy. As a
result, with reducing $\hbar \omega_0$ or increasing $B$, the total
angular momentum of two-electron increases, as previously discussed
by Wagner {\it et al}.~\cite{Wagner92} for gated 2D QDs.

Figure~\ref{phase}(c) shows the phase diagram of two interacting
electrons calculated by FCI method but with the vanishing spin Zeeman term, i.e. $g^\ast = 0$. This allows us to
distinguish the effects of spin Zeeman energy and the coulomb
interactions on the spin phase diagram of \fig{phase}(b), and also to
study the spin phases of QD made of a material with small $g^\ast$
such as GaAs. Without spin Zeeman splitting, the significant
features of \fig{phase}(c) are completely determined by the many-body
effects and geometry-engineered electronic structures of NWQDs.

In the $a>1$ regime, unlike the result shown in \fig{phase}(b), the $\rm{T_{L0}}$ phase disappears and naturally there is
no $\rm{S}$-$\rm{T_{L0}}$ transition observed. This is because the Coulomb
correlations that energetically favor spin-singlet state as mentioned previously, become dominant and compensate the negative exchange energy gained by the $\rm{T_{L0}}$ states.~\cite{Reimann02,Hanson07}
However, in the small $\hbar \omega_0$ regime, an additional singlet-triplet state oscillation with decreasing $\hbar \omega_0$  is observed.
Compared with Figure~\ref{phase}(b), the difference is the emergences of various singlet states between the triplet phases. This is due to the removal of spin Zeeman splittings, which energetically favor only the triplet states.  Such a singlet-triplet state oscillation is evidenced as a main feature of a flat 2D QD with small spin Zeeman effect, as shown both theoretically~\cite{Wagner92} and experimentally~\cite{Tarucha07} in the previous studies.

\section{Summary}

In conclusion, we present exact diagonalization studies of spin
phase transitions of two electrons confined in nanowire quantum dots
with highly tunable aspect ratio and external magnetic field.  A
configuration interaction theory based on a 3D parabolic model for
such three dimensionally confining QDs is developed, which provides
generalized explicit formulation of the Coulomb matrix elements and
allows for straightforward implementation of direct diagonalization.
The exact diagonalization study reveals fruitful features of spin ST
transitions with respect to the tunable geometric aspect ratio and
applied magnetic field.

For disk-like QDs, the ST transition behaviors may be dominated by
the spin Zeeman, the direct-Coulomb, and the exchange energies.
The pronounced Coulomb correlations are identified in rod-like QDs with
aspect ratio $a > 3$, which energetically favor singlet spin states
and yield the always non-zero critical magnetic fields of ST transitions.
The developed theory is further employed to study spin phase diagram
in the dimensional ``cross over'' regime from the 2D (disk-like) QDs
to finite 1D (rod-like) QDs. In the 2D disk-like QD regime, various
distinctive spin phases are emerged under the conditions of
appropriate lateral confinement strength and magnetic fields. In the
rod-like QD regime, switching the ST transitions is shown feasible
by controlling both lateral and/or longitudinal confinement
strength.

\section{Acknowledgement}

This work was financially supported by the National Science Council
in Taiwan through Contracts No. NSC-98-2112-M-009-011-MY2 (SJC) and
No. NSC97-2112-M-239-003-MY3 (CST). The authors are grateful to the
facilities supported by the National Center of Theoretical Sciences
in Hsinchu and the National Center for High-Performance Computing in
Taiwan.

\end{document}